\documentclass[twocolumn,showpacs,preprintnumbers,amsmath,amssymb]{revtex4}

\usepackage{graphicx}
\usepackage{dcolumn}
\usepackage{bm}

\begin{document}

\preprint{APS/123-QED}

\title{Direct observation of the decay of first excited  Hoyle state in $^{12}$C}

\author{T. K. Rana, S. Bhattacharya, C. Bhattacharya, S. Kundu, K. Banerjee, T. K. Ghosh,   G. Mukherjee, R. Pandey, M. Gohil, A. Dey,  J. K. Meena, G. Prajapati, P. Roy, H. Pai, M. Biswas}

\affiliation{%
Variable Energy Cyclotron Centre,  1/AF  Bidhan Nagar, Kolkata - 700 064, INDIA 
}%

\date{\today}

\begin{abstract}
An  excited  state of $^{12}$C having excitation energy E$_x \sim$~ 9.65~$\pm$~0.02 MeV and width (FWHM) $\sim$~607~$\pm$~55 keV, which decays to three $ \alpha $-particles via Hoyle state ($E_x \sim$~ 7.65 MeV), has been directly identified for the first time in the exclusive inelastic scattering of 60 MeV $^{4}$He on $^{12}$C, measured in coincidence with the recoiling   $^{12}$C$ ^* $ Hoyle state (decaying mostly as  $^{12}$C$ ^* $ $\rightarrow \  ^{8} $Be +   $ \alpha $  $\rightarrow \ \alpha + \alpha + \alpha$) by  event-by-event kinematic reconstruction of the completely detected (4$ \alpha $) events.  This state is likely to be a candidate for 2$_2^+$ first excited Hoyle state, the existence of which has recently been indirectly evidenced from the recent inclusive  inelastic scattering studies.   
\end{abstract}

\maketitle

 The famous Hoyle state \cite{hoyle}, the  $ 0^+_2 $ resonant excited state of $^{12}$C at an excitation energy of 7.654 MeV,  plays an important role  to understand  a variety of problems of nuclear astrophysics like elemental abundance in the universe  as well as the stellar nucleosynthesis process as a whole \cite{CN}. From nuclear structure point of view too, there are many unanswered questions regarding the configuration of this state; theoretically, it is conjectured as the lowest state corresponding to a different configuration ( member of either a $ \beta $ band of the three $ \alpha $ molecule-like structure \cite{kami,ueg,clus3}, or a Bose-Einstein condensate-like structure (BEC) \cite{cond}) originating from 3-$ \alpha $ clustering in $^{12}$C, and the standard shell-model approaches, even the advanced no-core calculations failed to reproduce  the state \cite{nocore}. 
 
According to the above models, the  3-$ \alpha $ cluster configuration of the Hoyle state ($ 0^+_2 $ at 7.654 MeV should also have higher excited states; the lowest excited state has been predicted to be a $ 2^+ $ state at excitation energy E$_x \sim$ 10 MeV \cite{kami,ueg,cond}. This $ 2^+_2 $ state is strongly coupled to the $ 0^+_2 $ Hoyle state and is likely to decay mostly via Hoyle state. However, in spite of vigorous experimental efforts in the recent years, there is still now  no conclusive evidence so far. In inelastic proton scattering $^{12}$C($p, \ p^\prime$) experiments,  small angle angular distribution measurement near the diffractive minimum of the broad $ 0^+ $ background has indicated the presence of a possible $ 2^+ $ state at 9.6(1) MeV of width $ \sim $600(100) keV \cite{fr1,fr2}. Recent inelastic $\alpha$-scattering angular distribution studies also indicated the presence of a $ 2^+ $ state at 9.84~$ \pm $~0.06 MeV of width 1.01~$ \pm $~0.15 MeV \cite{ito}. On the other hand, the study of $^{12}$C$ ^* $, produced in the $ \beta $-decay of $^{12}$N and $^{12}$B, decaying into 3$ \alpha $ continuum  has  however, not found clear evidence about the existence and nature of the  $ 2^+ $ and $ 0^+ $ states  at excitations $ \sim $10-12 MeV \cite{hyl1,hyl}. On the other hand, the $ \gamma $ induced $^{12}$C dissociation $^{12}C(\gamma, 3 \alpha)$ studies have also indicated the presence of a $ 2^+ $ state below 10 MeV \cite{gai}. 

It is thus clear that even though there are definite indications about the existence of the elusive $ 2^+_2 $ state, the first excited state of the Hoyle state, clear identification of the state is still missing. Assuming the 3-$ \alpha $ cluster configuration of $^{12}$C, the isoscalar (IS) transition rates to various excited states have been calculated by Khoa et al. \cite{khoa}, who have shown that the excited states of the Hoyle state band ( $ 0^+_2, 2^+_2, 4^+_2, ... $ should predominantly decay by E2 transitions to the ground state $ 0^+_2 $, the Hoyle state, which will then decay predominantly via two step process: $^{12}$C$^{*}$ $\rightarrow$ $^{8}$Be + $\alpha$ $\rightarrow$ $\alpha$  + $\alpha$  + $\alpha$, with a small percentage ($ \lesssim 5$\%  of direct $^{12}$C$^{*}$ $\rightarrow  3\alpha$ decay. So, complete kinematical measurement of all outgoing particles in each event may be helpful in reconstructing the the events originating from the decay of the excited states of the Hoyle state using missing energy (due to the emission of $ \gamma $-ray) criterion. However, a recent experiment performed in this line \cite{kirs}, where complete kinematical measurement of all outgoing particles emitted from the reactions $^{10}$B($^3$He, $p\alpha\alpha\alpha$) at 4.9 MeV and $^{11}$B($^3$He, $d\alpha\alpha\alpha$) at 8.5 MeV has been done, did not bring out any signature of the possible existence of the $ 2^+_2 $ state. Since the spin and isospin zero  $\alpha$-particle is a very good projectile for the excitation of the nuclear IS states \cite{khoa}, we studied the decay of $^{12}$C$^{*}$ into 3-body final states (3$\alpha$) using inelastic $\alpha$ scattering from $^{12}$C target to study the excited states of $^{12}$C which are predominantly excitable through isoscalar transitions in general, and to look for a cleaner signature of the elusive $ 2^+_2 $ state in particular. 
In this letter we report, for the first time,  a complete kinematic measurement of the inelastic $\alpha$-particles emitted in the $^{12}$C($\alpha$, $\alpha^\prime)$ reaction in coincidence with  the  decay of Hoyle state. The present study clearly  demonstrates the presence of an excited state of $^{12}$C at excitation energy of 9.65~$\pm$~0.02 MeV energy and width (FWHM) 607~$\pm$~55 keV. Since the state is decaying via the $ 0^+_2 $Hoyle state, and no direct (3$\alpha$) decay of this new state has been observed, it may be taken to be the excited state of the Hoyle state band.

The experiment was performed  at the Variable Energy Cyclotron Centre, Kolkata, using 60 MeV $ ^{4} $He ion beam from the K130 cyclotron on  $^{12}$C target (self supported, thickness $ \sim $90 $\mu$g/cm$ ^{2} $). The Hoyle state (and other excited states) in $^{12}$C nuclei were produced through inelastic scattering of $ ^{4} $He from $^{12}$C. The $\alpha$-particles emitted in the decay of Hoyle state as well as the inelastically scattered $ ^{4} $He have been detected in coincidence using two 3-element telescopes. The telescopes consisted of a 50$\mu$m $\Delta$E single-sided silicon strip detector (16 strips, each  of dimension 50mm $ \times $3mm), 500$\mu$m  $\Delta$E/ E double-sided silicon strip detector (16 strips (each 50mm $ \times $3mm)  per side in mutually orthogonal directions) and backed by four CsI(Tl) crystals (thickness 6 cm). The two telescopes were placed at kinematically correlated angles for  coincident detection of inelastically scattered $ ^{4} $He in the backward angle telescope (covering the angular range of 104$ ^{\circ} $ - 128$ ^{\circ} $) and the three $\alpha$-particles, originating from the decay of the Hoyle state of the recoiling  $^{12}$C$ ^{*} $, at the forward angle telescope (covering the angular range of 14.3$ ^{\circ} $ - 37.7$ ^{\circ} $).  All strips and the CsI(Tl) detectors were read out individually using standard readout electronics. A VME-based online data acquisition system was used for the collection of data on event-by-event basis.

The aim of the experiment was to identify the (unbound) excited  states of $^{12}$C by exclusive complete kinematical  measurement of all outgoing particles; only completely detected events (events where all four $ \alpha $ particles, three from the decay of $^{12}$C$ ^{*} $, as well as the inelastically scattered one were detected separately) have been used for the present analysis to remove any ambiguity about the origin of the detected particles. The system $ ^{4} $He +  $^{12}$C was chosen for this purpose for its specific advantage regarding the detection of complete events, as it has only a few open reaction channels compared to other heavy ion induced reactions.   
One horizontal collimator (6 mm width) was placed in front of the backward telescope such that data taking was restricted to only a few ($ \sim $2) strips around the median plane. So, the corresponding coincident recoiling  $^{12}$C$ ^{*} $ nucleus in the forward telescope was also restricted around the  median plane; this helped to enhance the percentage of completely detected events (three decaying $\alpha$-particles  confined within the span of the forward telescope and detected) among the whole set of coincident events. Typical beam current used for the experiment was $ \sim $5-10  nA. In total, nearly 4000 completely detected events were collected in the present experiment which have been analysed further to extract the structure of the Hoyle state.

\begin{figure}
  \includegraphics[width=7cm,height=6cm]{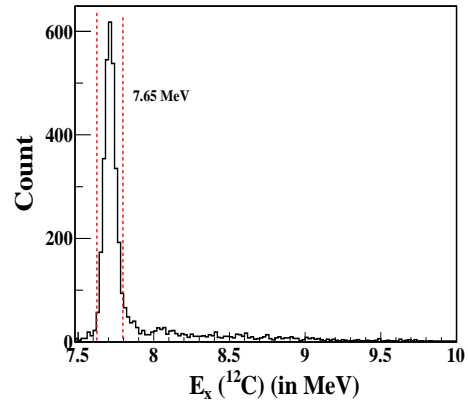}\\
  \caption{(color online) The excitation energy of recoiling $^{12}$C reconstructed from the three $\alpha$-particles emitted in the $^{12}$C$^{*}$ $\rightarrow  3\alpha$ decay. The red lines indicate the gate used for further analysis (see text).} 
  \label{fig0}
\end{figure}

The analysis of the data has been carried out in steps. In the first step, the energies and momenta of the three $\alpha$-particles detected in the forward telescope (completely detected event) have been used to reconstruct the excitation energy of the recoiling $^{12}$C$ ^{*} $, which has been displayed in Fig.~{\ref{fig0}}. It is seen that the excitation energy has only one prominent peak at E$ _x $($^{12}$C)$ \sim $ 7.65 MeV, which corresponds to the Hoyle state.

\begin{figure}[h]
  \includegraphics[width=7cm,height=6cm]{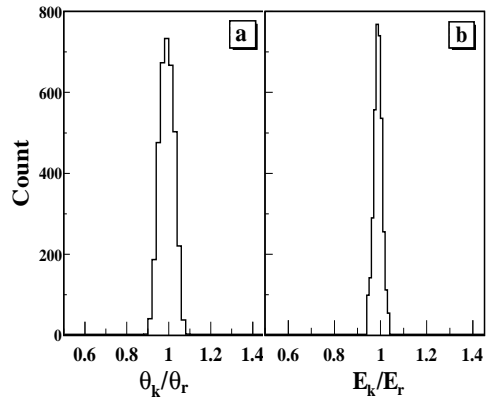}\\
  \caption{Comparison of the  (a) emission angle, and, (b) kinetic energy of the recoiling $^{12}$C$ ^{*} $ estimated by binary kinematics ($ \theta_k $, $E_k$) and  kinematic reconstruction ($ \theta_r $, $E_r$) methods (see text).} 
  \label{fig1}
\end{figure}

In the next step,  the positions, energies of the three identified $\alpha$-particles have been used to reconstruct the recoiling energy, position of $^{12}$C$ ^{*} $ nucleus ($E_r$, $ \theta_r $) and these values  have then been cross-checked with the  energy, position of $^{12}$C$ ^{*} $  extracted from the backward angle inelastic $\alpha$-particle data using binary kinematics ($E_k$, $ \theta_k $). The comparison is shown in Fig.~{\ref{fig1}}, which clearly demonstrates the consistency and precision of the reconstruction technique. 
 
\begin{figure}[h]
  \includegraphics[width=7cm,height=6cm]{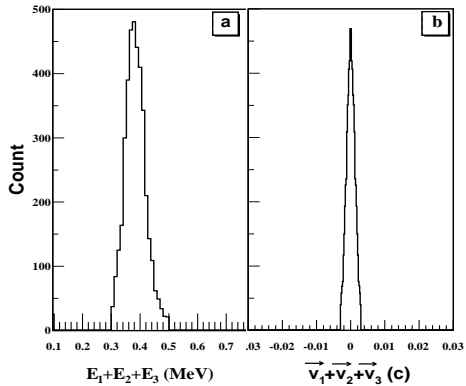}
  \caption{Consistency check of the transformation to the rest frame of $^{12}$C$ ^{*} $}
  \label{fig2}
\end{figure}

In order to study the decay of the Hoyle state further, the data have been transformed from the laboratory frame to the rest frame of $^{12}$C$ ^{*} $, which is characterised by the following conditions,

\begin{eqnarray}
\vec{v_1}+ \vec{v_2} + \vec{v_3}& = &0 \\
E_1 + E_2 + E_3& =& E_{breakup},
\label{eq1} 
\end{eqnarray}

where, $\vec{v_i} $ and $ E_i $, ($ i = 1 - 3 $) are the velocities and kinetic energies of the three $\alpha$-particles in the rest frame of $^{12}$C$ ^{*} $ and $ E_{breakup} $ is the the difference of the Hoyle state energy and the 3$\alpha$-decay threshold energy (=380 keV). The transformed data have been found to satisfy both the conditions (see Fig.~\ref{fig2}).   

\begin{figure}
  \includegraphics[width=7cm,height=6cm]{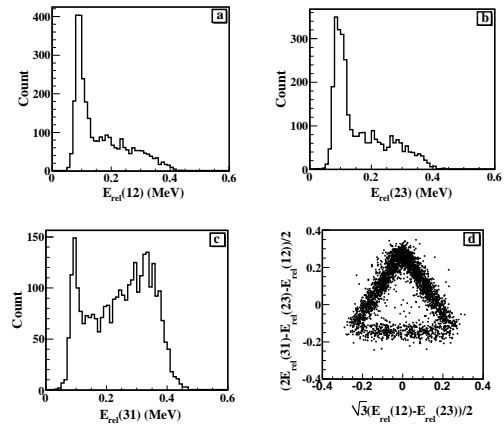}\\
  \caption{(a,b,c) Relative energy spectra for the three decay  $\alpha$-particles, and, (d) the Dalitz plot, for the decay of the Hoyle state.}
  \label{fig3}
\end{figure}

The nature of decay of the selected events have been  investigated further to check if it follows the known decay characteristics of the Hoyle state (predominantly sequential in nature: $^{12}$C$^{*}$ $\rightarrow$ $^{8}$Be + $\alpha$ $\rightarrow$ $\alpha$  + $\alpha$  + $\alpha$). The study   has been carried out using Dalitz plot technique \cite{dalitz}, utilising the relative energy spectra of the decay particles. The relative energy spectra and the corresponding Dalitz plots for the Hoyle state  have been shown in Fig.~\ref{fig3}. Here, the relative energy indices 1, 2 and 3 refer to the    particles emitted with highest, second highest and lowest energies, respectively. All relative energy spectra (Figs.~\ref{fig3}a-c) are found to be peaking sharply around $ \sim $90 keV, corresponding to the relative energy of the $^{8}$Be(g.s.)$\rightarrow$2$\alpha$ breakup.    The Dalitz plot (Fig.~\ref{fig3}d) has been generated using the Dalitz parameters $\sqrt{3}$(E$_{rel}$(12)-E$_{rel}$(23))/2 and (2E$_{rel}$(31) - E$_{rel}$(12)- E$_{rel}$(23))/2, where E$_{rel}$(ij) is the relative  energy between  i$^{th}$ and j$^{th}$  particle. The triangular locus in Fig.~\ref{fig3}d indicates that the decay is mostly  sequential in nature (sequential : $^{12}$C$^{*}$ $\rightarrow$ $^{8}$Be(g.s) + $\alpha$ $\rightarrow$ $\alpha$  + $\alpha$  + $\alpha$), in agreement with the earlier findings \cite{dsm}.   

\begin{figure}[h]
  \includegraphics[width=7cm,height=6cm]{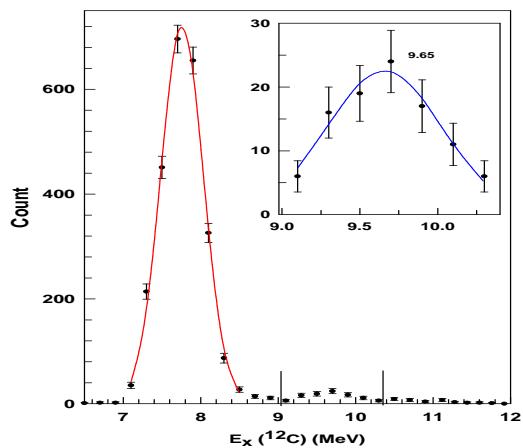}\\
  \caption{(color online) Excitation energy spectrum of $^{12}$C from inelastic $ \alpha $-scattering data gated with Hoyle state. The zone marked by the lines has been shown in the inset. The symbols represent the data and the lines are the corresponding fits to the data (see text). }
  \label{fig4}
\end{figure}

From the above, it is evident that the 3$\alpha$ reconstructed 7.65 MeV state in Fig.~\ref{fig0} is the Hoyle state. In the next step, the $^{12}$C$^{*}$ excitation energy spectrum has been generated from  the inelastic $\alpha$-scattering data of the backward telescope in coincidence with a gate on the observed Hoyle state in Fig.~\ref{fig0}, which has been shown in Fig.~\ref{fig4}.

It is clearly seen in Fig.~\ref{fig4} that the $^{12}$C$^{*}$ excitation energy spectrum obtained from  the inelastic $\alpha$-scattering contains two peaks. The first peak, at 7.73~$ \pm $~0.09 MeV is the $ 0^+_2 $ Hoyle state.  In addition, there is a small peak also seen at excitation energy $\sim$9.65~$ \pm $~0.02 MeV (see inset of Fig.~\ref{fig4}) . The  width of the Hoyle state, which is actually negligible (see Fig.~\ref{fig0}), appears to be quite broad in the inelastic scattering spectrum (Fig.~\ref{fig4}). This is due to the fact that the inelastic scattering spectrum has been generated by summing over a large solid angle to extract statistically significant information about the excited state; the observed broadening is therefore of kinematic origin associated with the total angular coverage. The excited state observed at 9.65~$ \pm $~0.02 MeV is having a large intrinsic width, which is estimated to be $\sim$607~$ \pm $~55 keV, obtained after correcting for the kinematic broadening. Since this state has been seen in coincidence with the Hoyle state, it is likely to be an excited state of $^{12}$C, which is decaying via Hoyle state. However, since there are quite a few states at around the same energy, it warrants a thorough introspection of the whole scenario.   

Most prominent and known states around this energy are, a narrow 9.64 MeV 3$^-_1$ state (width $ \sim $ 34 keV ), a broad $ \sim $10.3 MeV $ 0^+_3 $ state (width $ \sim $ 2.7 - 3.0 MeV \cite{ito,ajz}), a broad 10.84 MeV  $ 1^- $ state (width $ \sim $ 315 keV \cite{ajz} ). The 9.64 MeV 3$^-$ state is known to have a $\alpha$ decay branch which decays to $^{8}$Be(g.s.) (97.2\%) and $^{8}$Be(2.7$^+$\%)\cite{fr3}. The  inelastic $\alpha$ energy corresponding to the  $^{8}$Be(g.s) decay branch of 9.64 MeV 3$^-$ state cannot interfere with the present observation as the energy of the 3$\alpha$ reconstructed spectrum will be around 9.64 MeV, well outside the gate (see Fig.~\ref{fig0}). On the other hand,  $^{8}$Be(2$^+$) decays almost entirely ($ \approx $100\% ) into 2$\alpha$ \cite{nndc}; in such case, the 3$\alpha$ reconstructed spectrum will again be completely out of the gate used (Fig.~\ref{fig0}). Even  taking into consideration  those rare events ($ \ll $ 1\% of the total events) where $^{8}$Be(2$^+$) $\rightarrow$ $^{8}$Be(g.s.) through the emission of 3.03 MeV $\gamma$-ray, the three decay $\alpha$-particles   will not satisfy the kinematic conditions of the Hoyle state decay, and thus will automatically be rejected. Same argument is valid for all higher excited states, so far  as the decay via  $^{8}$Be + $\alpha$ channel is concerned.

\begin{figure}[h]
  \includegraphics[width=7cm,height=6cm]{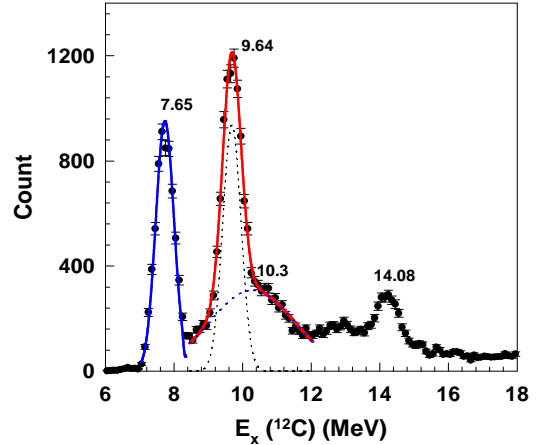}\\
  \caption{(color online) Excitation energy spectrum of $^{12}$C from inelastic $ \alpha $-scattering data gated with $^8$Be. The symbols represent the data and the lines are the corresponding fits to the data (see text).}
  \label{fig5}
\end{figure}

That the exclusive  measurement of the inelastic $^4$He spectrum in coincidence with the fully reconstructed Hoyle state has made all the difference is further evident from Fig.~\ref{fig5}, which displays the spectrum of inelastically scattered $^4$He in coincidence with the detected 2$\alpha$ (in forward telescope) events  (not only kinematically complete events as considered earlier), which follow the sequential decay (via  $^{8}$Be~+~$\alpha$) route. Here many prominent unbound  states of $^{12}$C$^{*}$ are clearly identified. It is observed in Fig.~\ref{fig5} that, apart from the Hoyle state, there is a strong peak alongwith a broad bump near E$_x \sim$~9-11 MeV region.  Gaussian peak fitting analysis revealed the presence of two peaks in that region; a narrow peak at 9.68~$ \pm $~0.04 MeV (width $\sim$50~$ \pm $~21 keV) and a broad peak at 10.27~$ \pm $~0.05 MeV (width $\sim$2.85~$ \pm $~0.13 MeV). The first peak is likely to be the 9.63 MeV, 3$^-$ state, whereas the broad peak may be identified as the 10.3 MeV,$ 0^+_3 $ state. The $ 1^- $ state at 10.84 MeV, which was quite prominent in  $^{12}$C($p, \ p^\prime$) experiment \cite{fr2}, is not found to be significant here; however the present observation is in agreement with the previous $^{12}$C($\alpha$, $\alpha^\prime)$ result \cite{ito}, where the contribution of 10.84 MeV $ 1^- $ state was shown to be small compared to that of the broad $ 0^+ $ 
bump. It is thus clear from the above that unambiguous identification of the excited $ 2^+_2 $ state of the Hoyle state band is extremely difficult, if not impossible, in presence of the  large background of the strong 3$^-$,$ 0^+_3 $, and other neighbouring  states. However, the situation changes dramatically when one uses complete kinematical measurement; the background is fully eliminated and it is possible to extract  informations about very weak transitions as was shown in Fig.~\ref{fig4}.

The only way the states (9.64 MeV (3$^-$), 10.3 MeV  ($ 0^+_3 $ )  or 10.84 MeV  ($ 1^- $)) may contribute to the observed inelastically scattered state at 9.6 MeV is as follows; first it should decay to $ 0^+_2 $ Hoyle state through $\gamma$-decay and then the Hoyle state subsequently decays into 3$\alpha$ particles. In that case, the width of the observed state should be close to those of the above states. However, the extracted width ($\sim$607~$ \pm $~55 keV) of the observed 9.65~$ \pm $~0.02 MeV state  is nowhere close to the known widths of any of the above states ($ \sim $34 KeV for 3$^-$ state (9.64 MeV) \cite{ajz}, $ \sim $2.7 MeV for $ 0^+_3 $ state (10.3 MeV) \cite{ito,ajz},  or, $ \sim $315 KeV for 1$^-$ state (10.84 MeV) \cite{ajz}). Furthermore, recent theoretical calculation of E$\lambda$ transition strengths indicates that  the transitions 3$^-_1$~$\rightarrow$~$ 0^+_2 $, 0$^+_3$~$\rightarrow$~$ 0^+_2$  are very small; rather there is appreciable transition strengths for both 3$^-_1$, 0$^+_3$~$\rightarrow$~$ 2^+_2 $, the excited state of the Hoyle state band \cite{khoa}. This implies, at least qualitatively, that the decay of 3$^-_1$, 0$^+_3$ states, instead of interfering with the signature of $ 2^+_2 $ state decaying primarily via Hoyle state, may rather contribute to the yield of the $ 2^+_2 $ state.

It may therefore be concluded that the  $\sim$9.65~$ \pm $~0.02 MeV excited state of $^{12}$C   seen in the inelastic $\alpha$-scattering spectrum in coincidence with the Hoyle state reconstructed from  kinematically complete events, is most likely a new excited state decaying via Hoyle state. Exclusive coincidence measurement with Hoyle state and complete kinematic reconstruction helped to rule out the possibility of linking this state to other excited states  (e.g., 3$^-_1$, 0$^+_3$) at around this excitation energy. In absence of detailed angular distribution measurement, it is not possible to assign the spin, parity of this state. The energy ($\sim$9.65~$ \pm $~0.02 MeV) and width ($\sim$607~$ \pm $~55 keV) of this new state are quite close to the those predicted for the first excited $ 2^+_2 $ state of the Hoyle state band from the inelastic scattering angular distribution studies, i.e., energy $\sim$9.6~$ \pm $~0.1 MeV, width $\sim$600~$ \pm $~100 keV)\cite{fr1,fr2}, and energy $\sim$9.84~$ \pm $~0.06 MeV, width $\sim$1.01~$ \pm $~0.15 keV)\cite{ito}. Since the state has been directly observed for the first time decaying directly via Hoyle state,  this new state is a strong candidate for the $ 2^+_2 $ state of $^{12}$C. Detailed angular distribution of this state should be carried out for proper identification of this state.


The authors thank the cyclotron operating staff for smooth running of the machine during the experiment.

\end{document}